# Bio-inspired surface coil for small-bore MRI at 15.2 T

S. E. Solis-Najera[1], F. Vazquez[1], J. Lazovic[2], L. M. Zopf[2], R. Martin[1], L. Medina[1], O. Marrufo[3], A. O. Rodriguez[4]

[1]Department of Physics, Faculty of Sciences, UNAM, Mexico City 04510, [2]Biocenter Core Facilities, Vienna 1030, Austria, [3]Department of Neuroimage, INNN MVS, Mexico City 14269, Mexico. [4]Department of Electrical Engineering, UAM Iztapalapa, Mexico City 09340, Mexico.

## I. Introduction

The gains in sensitivity, resolution, and information content are highly nonlinear with the magnetic field strength. The ultrahigh field (UHF) MRI will enable fundamental investigations of important phenomena not accessible with the current technologies [1]. The installed base of preclinical MR imagers across the world demands RF coils with these characteristics for a number of applications. Specific examples include the development of RF coils for preclinical MRI applications with improved performance and low Specific Absorption Rate (SAR). This type of RF coils and coil arrays are usually in grand demand [2]. A review of RF coils commonly used in animal model MRI can be found in [3]. The geometry of the coil determines the important aspects of a good coil design: good uniformity and high Signal-to-Noise Ratio (SNR). We looked for a biological analogy to developed an improved surface coil to acquire MR images. Nature offers engineering solutions that can be exploited to improve the performance of devices, such as MRI RF coils. In this paper, we proposed a coil layout based on flower petal patterns for UHF MRI at 15.2 T. This particular configuration offers the possibility to use a different number of petals and shapes. The petal resonator coil [4] and the magnetron coil [5] have similar layouts, which have proved to outperform the popular circular coil.

## II. Method

The individual petals in the present design represent a way to increase the magnetic flux density generated by the coil. By doing so, we expected to improve the coil performance in both the SNR and the magnetic field uniformity. This necessarily implies that the number and shape of the petal play an important role. We decided to use circular-like petals to facilitate the construction and its analysis. A prototype of the bio-inspired surface coil was built, including six circular-type slots (0.45 cm dimeter) and total coil radius of 20 mm. The coil prototype was constructed using copper sheets laminated onto a nonconductive board. Fig. 1 shows a photograph of the coil prototype and electronic components and dimensions.

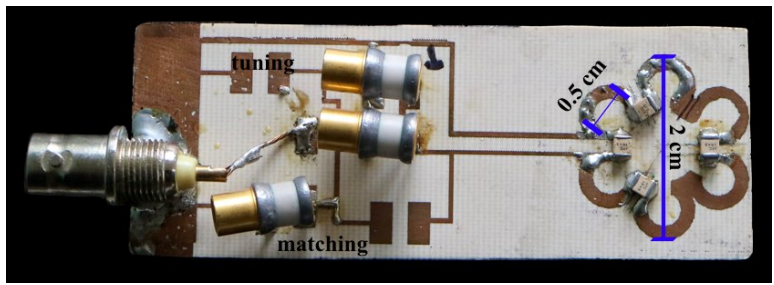

Figure 1. Photograph of the bio-inspired surface coil.

Tuning and matching capacitors (0–15 pF: Voltronics Co. Salisbury, MD, USA) were soldered directly onto the surface: two parallel ceramic capacitors (American Technical Ceramics, Huntington Station, NY,

USA) were placed as shown in Fig. 1. The coil prototype was then matched and tuned to 50 Ω and 650 MHz, respectively (proton frequency for 15.2 T). The signal received by the coil was conducted to an MR imager port via a 50-Ω-coaxial cable attached to the prototype coil. Fig. 2 shows the attenuation coefficient, $S_{11}$.

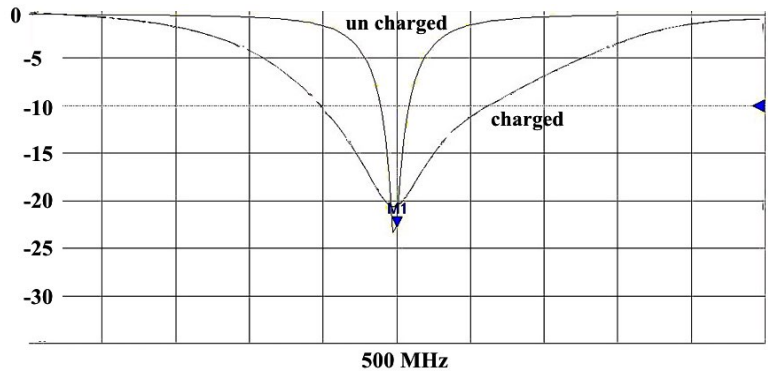

Figure 2. Attenuation coefficients for the charged and uncharged coil.

The quality factor (Q) of the coil was measured with a network analyzer (R&S-ZNB4, Rohde&Schwarz GmbH & Co., Munich, Germany) divided by 3-dB bandwidths, with a quarter-wavelength coaxial cable at the input of the coil. Coil performance was measured via the *Q*. Then, the loaded value was measured while the coil was placed on top of a saline solution-filled spherical phantom (20-mm radius). The coil design was operated in transceiver mode.

We performed phantom imaging experiments with a CuSO4 x 2H2O phantom in a 15.2T/11cm preclinical imager (BioSpec, Bruker Co, Ettlingen, Germany) using a standard gradient echo sequence (FLASH sequence): TE/TR = 1.6/100 ms, NEX = 2, Flip Angle = $25^0$, FOV = 18x18 mm², matrix size = 256x256, slice thickness = 1 mm, NEX = 4.

Fig. 3 shows a photograph of the small-bore MR imager used in all imaging experiments.

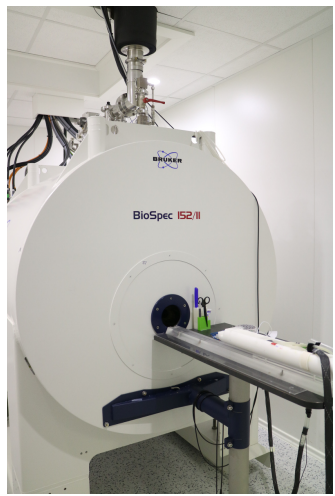

Figure 3. Photo of the preclinical MR imager for small animal models. This imager was equipped with a specific set of gradient coils for mouse's brain imaging.

The coil performance of the coil can be computed using the data of Fig. 2 for the unloaded and loaded coil. Additionally, the noise figure (NF) can be calculated using the following expression [5]:

$$NF = 10 \ln\left(\frac{Q_u}{Q_u - Q_L}\right) \quad (1)$$

where $Q_u$ and $Q_l$ represent the unloaded and loaded cases for a coil (see Fig. 2, too). Eq. (1) measures the reduction in SNR due to the losses in the coil only.

**III. Results and Discussion**

The attenuation coefficients showed the expected behaviour of a coil with a good performance. The quality factors of coil was approximately: $Q_u/Q_l$=21/13. These values compared well with those values abundantly reported in the literature. From Fig. 2, we can appreciate an RF penetration of around 22 dB, which indicates that good image quality can be obtained.

The *NF* was also computed for our coil using these *Q* values above giving 9.65. This bench test results and the *NF* show a reasonably good performance of the bio-inspired design. However, this value is greater than those reported at lower resonant frequencies. It is important to highlight that no data have been reported on the *NF* at the resonant frequency of 650 MHz or above. The number of both fixed-value capacitors and trimmers play an important role decreasing the SNR. This might explain the high *NF* value of this coil prototype. Future coil designs should consider reducing the number of capacitors to improve the performance of this type of coils.

Phantom images were acquired with our coil prototype and shown in Fig. 4. The black strips in the images represent the internal structures of the spherical phantom, and they are not image artifacts. These phantom images show no image degradation caused by the mutual inductance of the adjacent petals in both orientations.

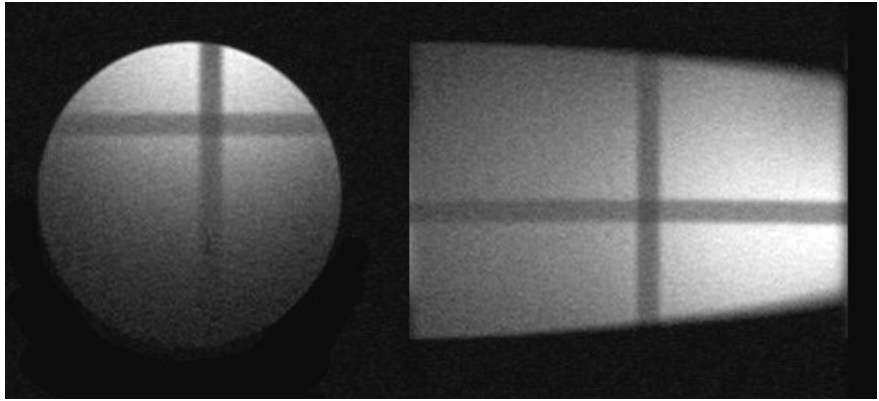

Figure 4. Phantom images acquired using standard gradient echo sequences.

The SNR was also calculated using the image data and giving 31.16. The results also confirmed the findings previously reported [2-4].

**IV. Conclusions**

This paper has presented a bio-inspired surface coil for small-bore MR imagers. Phantom images showed the coil viability and compatibility with conventional pulse sequences at 15.2 T. It still remains to investigate whether this bio-inspired surface coil is able to outperformed the popular circular coil. Other coil geometries can also be studied using other petal configurations to improve the coil performance for particular applications. This design may particularly prove to be optimal for mouse' brain imaging.

## V. Acknowledgements

We would like to thank CONACYT Mexico (grant number 112092) and, Innovation and Research/Technology Support Program (PAPIIT) UNAM (grant number IT 102116), and the Preclinical Imaging Facility at Vienna Biocenter Core Facilities, Austria. We extend much appreciation to Mr. Scott Ireland from Bruker, CO., for helping us to conduct this research.